\documentclass[preprint,12pt]{aastex}
\pdfoutput=1

\usepackage{url}
\newcommand{\degrees}{\mbox{$^\mathrm{o}$}}
\newcommand{\ms}{m~s$^{-1}$}
\newcommand{\kms}{km~s$^{-1}$}

\begin{document}

\title{Three Dimensional Modeling of Hot Jupiter Atmospheric Flows}

\author{Emily Rauscher \& Kristen Menou
  \\ \textit{Department of Astronomy, Columbia University,
  \\ 550 West 120th Street, New York, NY 10027, USA}}

\begin{abstract}
We present a three dimensional hot Jupiter model, extending from 200
bar to 1 mbar, using the Intermediate General Circulation Model from
the University of Reading.  Our horizontal spectral resolution is T31
(equivalent to a grid of 48$\times$96), with 33 logarithmically spaced
vertical levels.  A simplified (Newtonian) scheme is employed for the
radiative forcing.  We adopt a physical set up nearly identical to the model
of HD 209458b by \citet{CS05,CS06} to facilitate a direct model
inter-comparison.  Our results are broadly consistent with theirs but
significant differences also emerge. The atmospheric flow is
characterized by a super-rotating equatorial jet, transonic wind
speeds, and eastward advection of heat away from the dayside.  We
identify a dynamically-induced temperature inversion
(``stratosphere'') on the planetary dayside and find that temperatures
at the planetary limb differ systematically from local radiative
equilibrium values, a potential source of bias for transit
spectroscopic interpretations.  While our model atmosphere is quasi-identical to that of \citet{CS05,CS06} and we solve the same
meteorological equations, we use different algorithmic methods,
spectral-implicit vs. grid-explicit, which are known to yield fully
consistent results in the Earth modeling context.  The model
discrepancies identified here indicate that one or both numerical
methods do not faithfully capture all of the atmospheric dynamics at
work in the hot Jupiter context.  We highlight the emergence of a
shock-like feature in our model, much like that reported recently by
Showman et al. (2009), and suggest that improved representations of
energy conservation may be needed in hot Jupiter atmospheric models,
as emphasized by Goodman (2009).
\end{abstract}

\section{Introduction}

Hot Jupiter atmospheres challenge our understanding of meteorology,
introducing a new regime in which the atmospheric conditions are
significantly different from the more familiar solar system cases.
Common assumptions used in classic atmospheric dynamics may break down
and hot Jupiter models may require the inclusion of new physics.
These planets orbit their parent stars with periods of a few days and
are expected to be tidally locked, with permanent daysides subject to
intense stellar irradiation.  This strong, asymmetric heating is
clearly unique, but hot Jupiters also differ from their namesake
Jupiter in that their rotation periods are much longer (several days
instead of 10 hours), so that the size of dynamical atmospheric
structures should be much larger in scale \citep{SG02,Menou2003}.  For
a detailed discussion of the characteristics of this new regime see
the recent review by \citet{SMC08}.  This novel type of planetary
atmosphere challenges us to develop models that can correctly treat
the physics at work and thus be used to interpret recent
ground-breaking observations.

Aside from the inherently interesting properties of hot Jupiters, they
have also garnered attention because of impressive direct detections
of their atmospheres, which give theorists the opportunity to test
their models against reality.  The first observations of thermal
emission from the daysides of hot Jupiters
\citep{Charbonneau2005,Deming2005}, have been followed by multiwavelength measurements \citep{Charbonneau2008,Knutson2008,Knutson2009b}, the detection of a surprisingly bright planet \citep{Harrington2007}, observations of a Neptune-mass planet \citep{Deming2007,Demory2007}, and multi-epoch observations \citep{Agol2009}.
Orbital phase curves, which measure the variation in emission as different longitudes on the planet rotate into view, have been obtained
\citep{Harrington2006,Cowan2007,Knutson2007,Knutson2009a}.  This same observational technique has been used to witness the effect of flash-heating on a highly eccentric planet \citep{Laughlin2009}.  We are also able to determine the
composition and something of the vertical temperature profile of these
planets from transmission spectra \citep{Charbonneau2002,Tinetti2007,Pont2008,Sing2008,Swain2008b} as starlight
filters through the planet's atmosphere, and the planet's own emission
spectrum \citep{Grillmair2007,Richardson2007,Swain2008a}.  In light of
these increasingly detailed constraints, we must continue to improve
our atmospheric models, to reach a better understanding of the basic
physics at work and grow confidence in that we have correctly captured
the relevant processes.

Decades of work have gone into developing complex codes to model the
dynamics of the Earth's atmosphere, with extensions to include other
planetary bodies in our Solar System.  For the Earth, we have the
benefit of being able to directly compare the model results to the
circulation patterns observed \textit{in situ}.  So far for hot
Jupiters we only have basic, disk-integrated, usually single-epoch
observations, making it more difficult to test the reliability of
model results.  There exists a wide range of hot Jupiter models in the
literature, which employ a variety of methods and assumptions to
approach this novel regime
\citep{Cho2003,Cho2008,Burkert2005,CS05,CS06,Langton2007,DobbsDixon2008,S08,S09}. Comparing
the various models used against each other is important to test their
numerical and physical consistency.

Comparisons between atmospheric codes is standard practice in Earth
modeling.  \citet{HS94} presented a benchmark comparison between two
dynamical cores, i.e. codes solving the dynamical equations
independently of radiative forcing, for a simplified Earth
model.\footnote{For an additional comparison of three dynamical cores,
see for example
\url{http://www-personal.umich.edu/~cjablono/comparison.html}} Given
an identical, idealized physical set-up, they found excellent
quantitative agreement between the two numerical codes, suggesting
that each code was accurately solving the meteorological equations in
the Earth regime.  We would like to gain this level of confidence in
hot Jupiter circulation models.  Here we use a pseudo-spectral,
semi-implicit code to solve the same set of equations and atmospheric
conditions as done by \citet[][hereafter C\&S]{CS05,CS06} with the
grid-based, time explicit ARIES/GEOS dynamical core \citep{ST1995}.  The ARIES/GEOS code is identical to, and the IGCM conceptually similar to, the codes compared by \citet{HS94}, although
they are applied here to an entirely new circulation regime.  We pay
special attention to various details in our model results in order to
facilitate potential comparisons with other codes in the future.
Disparities in model results for an identical physical setup should
help us understand how the different algorithms deviate in the regime
of relevance to hot Jupiters and presumably lead us to the
identification of any additional physics that may be needed for the
accurate modeling of hot Jupiter atmospheres.

In \S\ref{sec:model} we introduce the code used and detail our
physical and numerical set up.  We present our results by first
showing horizontal structures at different vertical levels in
\S\ref{sec:horizontal} and then presenting vertical structure in
\S\ref{sec:vertical}.  We compare our results with those of C\&S in
\S\ref{sec:cs}, discuss issues of vertical exchange between the upper
and lower regions of the atmosphere in \S\ref{sec:vtransfer}, and in
\S\ref{sec:jump} provide a detailed discussion of the shock-like
feature found in our simulated flow.  We summarize our most
significant findings and conclude in \S\ref{sec:summary}.

\section{The Model} \label{sec:model}

We use the Intermediate General Circulation Model (IGCM) developed at
the University of Reading \citep{Hoskins1975}.  This is a well-tested
and accurate solver of the primitive equations of meteorology: the
three dimensional fluid equations applied to a shallow ideal gas
atmosphere in hydrostatic balance on a rotating sphere.  We use the
first version of the code
(IGCM1\footnote{\url{http://www.met.reading.ac.uk/~mike/dyn_models/igcm/}})
in which radiative forcing is implemented through a Newtonian
relaxation scheme.  This pseudo-spectral, semi-implicit code is
described in more detail in \citet{MR08}, but we repeat salient points
here.

The vertical coordinate used is $\sigma = p/p_s$, where $p_s$ is the
pressure at the bottom boundary and can vary horizontally.  In the run
presented here we obtain surface pressure variations of at most 0.5\%.
Thus each $\sigma$ level is fairly well represented by a constant
pressure, and for clarity of presentation we will refer to each level
by its average pressure.  Our figures will likewise use pressure as
the vertical coordinate.

The model solves the flow in a frame that is rotating with the bulk
planetary interior, which is assumed to be precisely synchronized with
the orbit. In the remainder of this paper, we refer to planet days as
equal to the rotational (and orbital) period of the planet.

The radiative forcing is treated through the method of Newtonian
relaxation.  The diabatic heating and cooling rate at every point in
the atmosphere is given by
\begin{equation}
\label{eq:q}
Q_T=\frac{T_{\mathrm{eq}}(\sigma,\lambda,\phi)-T}{\tau_{\mathrm{rad}}(\sigma)}
\end{equation}
\noindent where the local temperature, $T$, relaxes to the prescribed
equilibrium profile, $T_{\mathrm{eq}}$, on the radiative timescale
defined by $\tau_{\mathrm{rad}}$.  We describe our particular choice
of $T_{\mathrm{eq}}$ in the next section.

To model the effects of dissipation on small scales, hyperdissipation
acts upon the vertical component of the flow relative vorticity,
%$\vec{k} \cdot (\nabla \times \vec{v})$, 
the flow divergence, 
%$\nabla \cdot \vec{v}$, 
and the temperature fields as:
\begin{equation}
\label{eq:hyperdiss}
Q_{\mathrm{vor,div,}T}=-\nu_{\mathrm{diss}}\nabla^8[\mathrm{vor,div,}T]
\end{equation}
\noindent where the coefficient $\nu_{\mathrm{diss}}$ ($=8.54 \times
10^{47}$~m$^8$~s$^{-1}$ in the present model) is chosen so that the
smallest resolved structures are diffused in a small fraction of a
planet day.

There are several key distinctions between the work done here and in
\citet{MR08}.  We have significantly extended our vertical domain,
using logarithmically spaced levels to model an atmosphere from 1 mbar
to 220 bar.  In this paper the radiative timescale is not constant,
but increases with depth in the atmosphere.  We have also chosen a
relaxation temperature profile (described in detail below) with the
strongest day-night temperature differences at the top of the
atmosphere.  The significant variations with height introduced in this
deep model apparently prevent the same barotropic (vertically aligned)
behavior as reported in \citet{MR08}.

\subsection{Parameter choices} \label{sec:params}

We have chosen physical and model parameters so as to match the model
in C\&S as closely as possible, which is set up for the hot Jupiter HD
209458b.  Specifically, we adopt the same values for the planetary
radius, gravitational acceleration, and rotation rate: $R_p=$
9.44$\times 10^7$ m, $g=$ 9.42 m s$^{-2}$, and $\Omega_p=$ 2.06$\times
10^{-5}$ rad s$^{-1}$.  We use the same specific gas constant,
$\mathcal{R}=4593$ J kg$^{-1}$ K$^{-1}$, and
$\kappa=\mathcal{R}/c_p=0.321$.  The horizontal resolution in C\&S is
approximately [5\degrees, 4\degrees] in longitude and latitude.  We
run at T31 resolution, which corresponds to a resolution slightly
under 4\degrees~in both dimensions.  The model in C\&S spans a range
from 1 mbar to 3 kbar with 40 logarithmically spaced vertical levels.
We use the same vertical resolution, but only model down to 220 bar
(using 33 levels) so that our bottom boundary is safely above the
transition to the interior adiabat at $\sim$1 kbar \citep{Iro05} since
this version of our IGCM code does not include any special treatment
for convective columns.

As in C\&S we use the work by \citet{Iro05} to set the radiative
forcing scheme.  It is assumed that by 10 bar all of the incident
stellar photons will have been absorbed, and this separates the
atmosphere into radiatively ``active'' layers above 10 bar and
``inert'' layers below.  The active layers are heated with radiative
timescales ($\tau_{\mathrm{rad}}(\sigma)$ in Equation~\ref{eq:q})
taken from Figure~4 of \citet{Iro05}, converting to our vertical coordinate $\sigma$ by assuming a constant $p_s=220$ bar.  The inert layers experience no
diabatic heating or cooling, which we impose by setting
$\tau_{\mathrm{rad}}(P \ge 10 \mathrm{bar})\rightarrow \infty$.

The Newtonian relaxation temperature profile is set so that at each
level the nightside temperature is constant and the dayside
temperature decreases as $(\cos \alpha)^{1/4}$ where $\alpha$ is the
angle away from the substellar point \citep[see][Equation 2]{CS05}.
The temperature difference between the substellar point and the
nightside is set to 1000 K above 100 mbar and decreases
logarithmically with pressure down to 530 K at 10 bar.  The nightside
temperature is then chosen such that the averaged $T^4$ will equal
$T_{\mathrm{Iro}}^4$ at that pressure, where $T_{\mathrm{Iro}}(\sigma)$ is
taken from Figure 1 of \citet{Iro05}, using the model with an internal
temperature of 100 K, and converting from pressure to $\sigma$ by assuming a constant $p_s=220$ bar.  Figure 1 of \citet{CS06} shows the
corresponding substellar and nightside relaxation temperature profiles
(see also our Figure~\ref{fig:tprof} below).

The initial temperature field is set with each level at a constant
temperature, equal to the nightside temperature from the relaxation
profile for layers above 10 bar, or the $T_{\mathrm{Iro}}$ value for
layers below 10 bar.  Initially there are no winds.  To break initial
symmetry we include noise as small amplitude random perturbations in
the surface pressure field.

We note that our set-up results in discontinuities at 10 bar, both in the radiative timescales and the initial vertical temperature profile.  Due to the very weak radiative forcing in the deep atmosphere, these are effectively small discontinuities and we do not expect them to strongly influence the resulting flow.  We return to discuss this issue in Section 4.2.

We emphasize that we have set up the simulation so that it matches the
work of C\&S as closely as possible.  The main differences are: \emph{1)}
our model extends down to 220 bar while theirs includes additional
layers down to 3 kbar, \emph{2)} we have introduced a small amount of
initial noise, \emph{3)} our horizontal resolution is slightly finer,
and \emph{4)} by nature of being a spectral code, instead of
grid-based, we do not have any special treatment at the poles.  There
are additional differences introduced in the algorithmic methods each
code uses to solve the same set of equations and this is exactly what
we are testing in this work.

\section{Results} \label{sec:results}

Our results are broadly consistent with C\&S, but we also identify
important differences in the flow and temperature patterns.  This
demonstrates that the methods by which our codes solve the same set of
equations deviate at a significant level in the hot Jupiter regime.
We present our results in the following sub-sections and leave a
discussion of the implications until \S\ref{sec:discussion}, with a
careful comparison between our results and those of C\&S detailed in
\S\ref{sec:cs}.

Within one planet day winds at the top levels have reached supersonic
speeds, but the initial ramp-up period, during which kinetic energy
increasing at all levels, lasts for $\sim$450 planet days.  Due to the
shorter radiative timescales, the upper layers reach a statistically
stationary state sooner, but by $\sim$600 planet days all of the
layers down to 3 bar have reached a stationary state.  There is a very
low level of variability, with temperatures and wind speeds varying by
$\sim$5\% in the upper layers over the last 500 planet days of the
run.  Like C\&S, we focus on the final snapshot of the run, at 1450
planet days.

We also tested the effects of several modeling choices on our final
results.  A run without initial noise resulted in a very similar flow.
The absence of inert layers (only including layers above 10 bar) did
not significantly affect the upper atmospheric flow. Finally, a
start-up period during which we gradually imposed the radiative
forcing did not significantly alter the end behavior.

\subsection{Horizontal structure} \label{sec:horizontal}

\begin{figure}[ht]
\begin{center}
\includegraphics[width=0.7\textwidth]{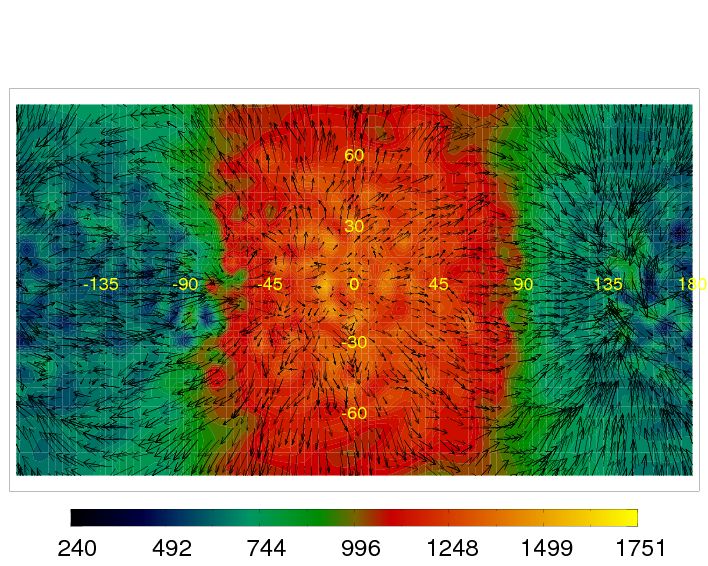}
\includegraphics[width=0.7\textwidth]{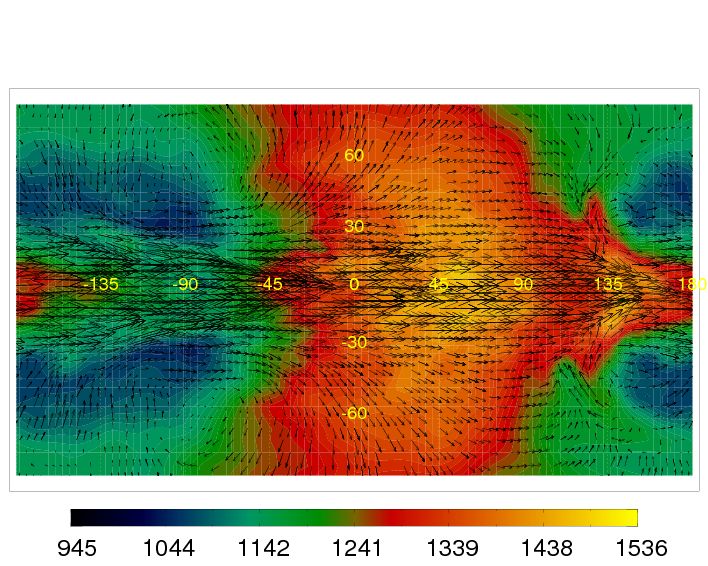}
\caption{Temperature (in K) and wind maps at pressure levels of 2.5
mbar (\emph{top}) and 220 mbar (\emph{bottom}) in our hot Jupiter
model.  Each map is shown in Miller cylindrical projection, with the
substellar point located at (0,0).  The maximum wind speeds are 10
(top) and 4 \kms (bottom).  High in the atmosphere, winds are very
strong and directed away from the dayside, but the temperature pattern
remains strongly determined by insolation.  Somewhat deeper in the
atmosphere, the winds become more efficient at advecting hot gas away
from the dayside before it cools significantly. A strong
super-rotating equatorial jet develops.}
\label{fig:levels1}
\end{center}
\end{figure}

High in the atmosphere, there is a strong day-night temperature
difference and wind speeds of several \kms~blow from day to night
across the east terminator and over the poles (see
Figure~\ref{fig:levels1}, top panel, at $\sim$2.5 mbar).  As a rough
measure of how quickly winds can move hot air from the dayside to the
nightside, we calculate the advective timescale,
$\tau_{\mathrm{adv}}=(\pi R_p)/[u]_{\mathrm{eqtr}}$, where $R_p$ is
the planet radius and $[u]_{\mathrm{eqtr}}$ is the zonal average of
zonal wind at the equator.  This advective timescale for east-west
flow along the equator is generally faster than for the north-south
flow.  Above $\sim$200 mbar, the radiative timescales are less than
the advective timescales and the temperature pattern is thus largely
determined by the insolation, despite the supersonic
\kms~winds.\footnote{Note that this argument neglects the possible
role of vertical transport.}

As the atmosphere transitions to having longitudinal advective
timescales shorter than the radiative ones, there is significant
advection of heated gas away from the substellar point, by
$\sim$45\degrees~to the east at $\sim$200 mbar.  The winds are still
supersonic, but now the flow is dominated by a super-rotating
(eastward) equatorial jet, approximately extending from 30\degrees~N
to 30\degrees~S in latitude (Figure~\ref{fig:levels1}, bottom panel,
at $\sim$220 mbar).  At high latitudes there are quasi-stationary
vortices at the west terminator.  Such vortices also exist in the
shallower model of \citet{MR08}, although at a longitude of
-135\degrees~rather than -90\degrees~like here.  There is also a
strongly convergent flow feature at a longitude of +135\degrees~that
we discuss in detail in \S\ref{sec:jump}.

\begin{figure}[ht]
\begin{center}
\includegraphics[width=0.7\textwidth]{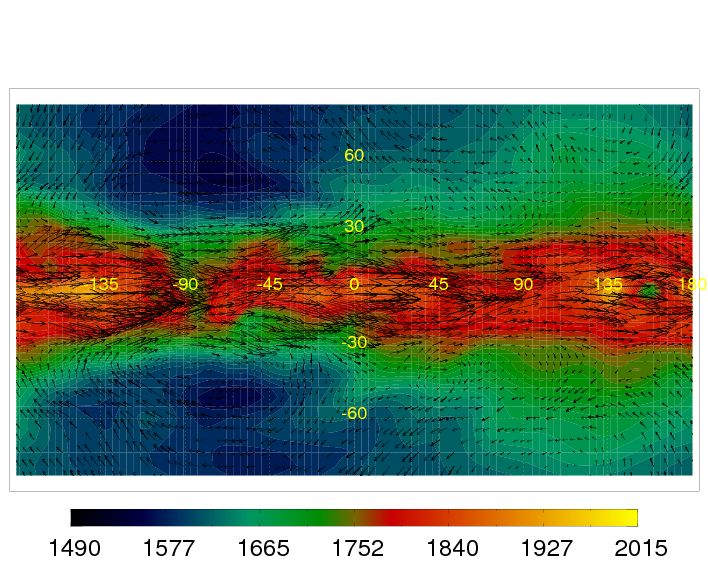}
\includegraphics[width=0.7\textwidth]{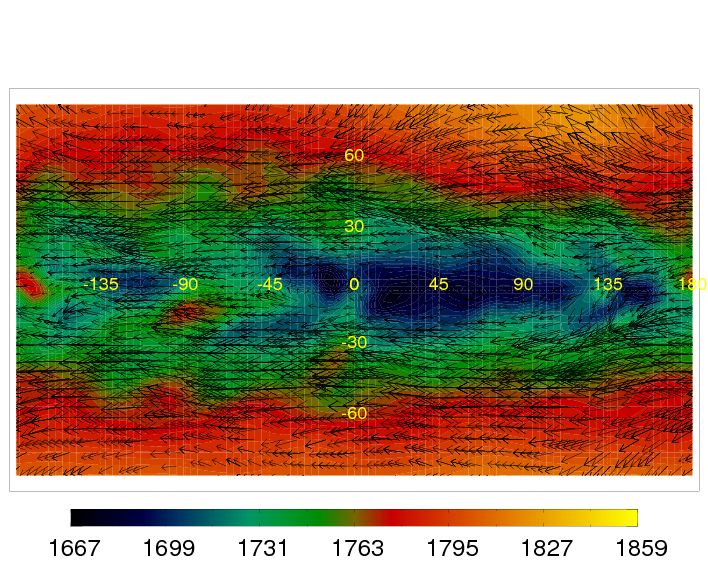}
\caption{Temperature (in K) and wind maps, continued from
Figure~\ref{fig:levels1}, at pressure levels of 4.4 bar (\emph{top})
and 20 bar (\emph{bottom}) in our hot Jupiter model.  Maximum wind
speeds are 1 (top) and 0.2 \kms (bottom). At these deeper levels, the
winds have largely homogenized the longitudinal temperature structure,
although some latitudinal gradient remains.  The super-rotating
equatorial jet descends through much of the atmosphere, down to
approximately 7 bar.}
\label{fig:levels2}
\end{center}
\end{figure}

Below this level the advective timescales are much shorter than the
radiative forcing times, the winds become subsonic, and the
temperature pattern is more longitudinally homogenized.  The
equatorial jet, which has extended down through much of the
atmosphere, ends at $\sim$4.4 bar (Figure~\ref{fig:levels2}, top
panel).  At this pressure level, vortices are found on the edges of
the jet, as well as around the poles.

Below 10 bar the atmosphere is no longer heated by the stellar
irradiation and the maximum wind speeds drop to several hundred \ms.
The 20 bar level (Figure~\ref{fig:levels2}, bottom panel) is located
below the main super-rotating jet (seen in Figure~\ref{fig:levels1},
bottom panel, and Figure~\ref{fig:levels2}, top panel) but also above
another deeper, weaker, super-rotating jet (see Figure~\ref{fig:uz}
below).  The equatorial flow at 20 bar is retrograde (westward) and
temperature increases with latitude, as expected from geostrophic
balance\footnote{Geostrophic balance in rotating stratified flows
corresponds to the Coriolis force being equal to and opposite the
pressure gradient.} with the anticyclonic flows around each pole.

\subsection{Vertical structure} \label{sec:vertical}

\subsubsection{Wind profiles}

\begin{figure}[ht]
\begin{center}
\includegraphics[width=0.6\textwidth]{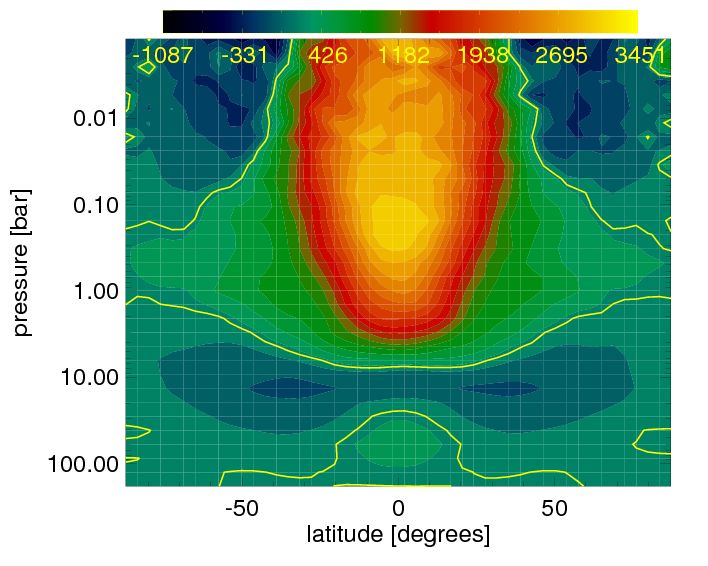}
\caption{Zonal average of the zonal wind in \ms, as a function of
latitude and depth in the atmosphere.  Yellow lines separate regions
of positive (eastward) flow and negative (westward) flow.  A strong
super-rotating equatorial jet descends through much of the atmosphere.
At the equator the maximum (eastward) zonal-average wind speed occurs
where the radiative and advective timescales are comparable. Deep in
the atmosphere, the minimum (peak westward) zonal-average wind speed
occurs at the transition between radiatively forced and inert layers.}
\label{fig:uz}
\end{center}
\end{figure}

For another view of how the flow structure changes throughout the
atmosphere, we plot the zonal average of the zonal wind as a function
of pressure and latitude, calculated as
\begin{equation}
[u(P,\phi)] \equiv \frac{1}{2 \pi} \int_0^{2\pi}u(P,\lambda,\phi)\ d\lambda
\end{equation}
\noindent where $u$ is the zonal (east-west) wind, $P$ is pressure,
and $\phi$ and $\lambda$ are latitude and longitude.
Figure~\ref{fig:uz} reveals the strong super-rotating jet that extends
throughout most of the heated atmosphere (above 10 bar).  The peak
zonal wind speed in the jet occurs at the depth where the radiative
and longitudinal advective timescales are roughly equal.  The short
radiative times in the upper atmosphere drive fast winds, but since
the flow there is directed from day- to night-side across all regions
of the terminator, the westward flow averages against the eastward
flow.  Thus the peak in the zonal wind speed occurs at a point of
balance between absolute wind strength and zonal wind coherence, which
is more clearly established at lower levels, where the flow is
dominated by the eastward jet.

As shown in Fig.~\ref{fig:uz}, the strongest westward flow at the
equator occurs around 10 bar, as the atmosphere transitions from
active to inert regions.  The inert lower layers also develop a
super-rotating jet, but it is much weaker than the one in the active
layers.  Although the mass of the atmosphere is dominated by the
lowest levels\footnote{A mass element is defined by $dm = \rho \ dx\
dy\ dz = - (1/g)\ dA\ dP$, where $\rho$ is the density, $g$ is the
gravitational acceleration, and $dA$ is a horizontal surface element.
Gravity is assumed constant throughout the model and our levels are
logarithmically spaced in pressure.}, the region between 80 mbar - 6
bar contains $\sim 2/3$ of the total kinetic energy of the atmosphere,
as we shall now see.

\begin{figure}[ht]
\begin{center}
\includegraphics[width=0.6\textwidth]{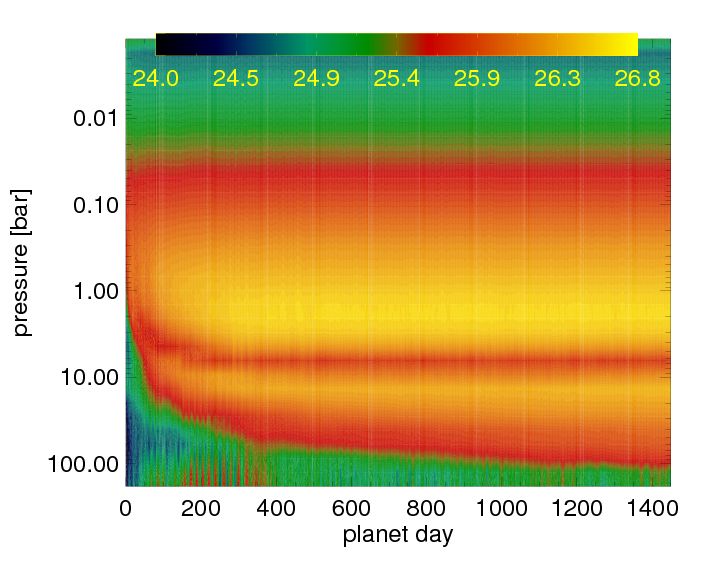}
\caption{Logarithm (log$_{10}$) of the total wind kinetic energy (in
joules) in each layer of the model atmosphere as a function of
atmospheric depth and planet day in the run.  The kinetic energy at
each level is calculated as the horizontal integral of $\rho v_H^2 /
2$, where $\rho$ is the density and $v_H$ is the horizontal wind
speed.  The atmosphere starts at rest but it quickly develops a
transonic flow.  The equatorial jet shown in Figure~\ref{fig:uz}
contains most of the kinetic energy of the atmosphere.  The inert
layers below 10 bar slowly gain kinetic energy by exchange with the
radiatively forced layers above, and they have yet to reach a
stationary state by the end of the run.}
\label{fig:ke}
\end{center}
\end{figure}

We plot the wind kinetic energy of each atmospheric layer as a
function of time for the length of our run in Figure~\ref{fig:ke} to
characterize the development of the modeled flow.  The atmosphere
begins at rest, but by planet day 1 supersonic winds have developed in
layers above 150 mbar.  By the end of the run the peak kinetic energy
is located at $P \sim 2$ bar, but it takes about 300 planet days for
these levels to fully adjust to the radiative forcing, where the local
radiative timescales are relatively long ($\approx$ 5 planet days).

Below 10 bar, the atmosphere is not being driven by direct radiative
forcing, but instead by transfer of heat and momentum from layers
higher up.  By the end of the run the layers below 10 bar have
horizontal temperature variations of less than 100 K from their
average values of $\sim$1800 K.  Since these are small temperature
differences, it seems that the main driver of the flow in the lower
atmosphere would be momentum transport from above.  This transport to
deeper levels continues throughout the entire run and there is
evidence that the lower levels have not reached a fully stationary
state by the end of our run.  We leave further discussion of concerns
related to this issue until \S\ref{sec:vtransfer}.

Lastly, we note the presence of some amount of horizontal kinetic
energy associated with layers just above the model bottom boundary,
especially early on, showing a quasi-oscillatory behavior. A run
without initial surface pressure noise shows a similar behavior. We
have not been able to unambiguously identify the origin of this flow
feature, which may be related to some form of vertical propagation
from the upper layers and subsequent interaction with the bottom
boundary.

\subsubsection{Temperature profiles}

\begin{figure}[ht]
\begin{center}
\includegraphics[width=0.6\textwidth]{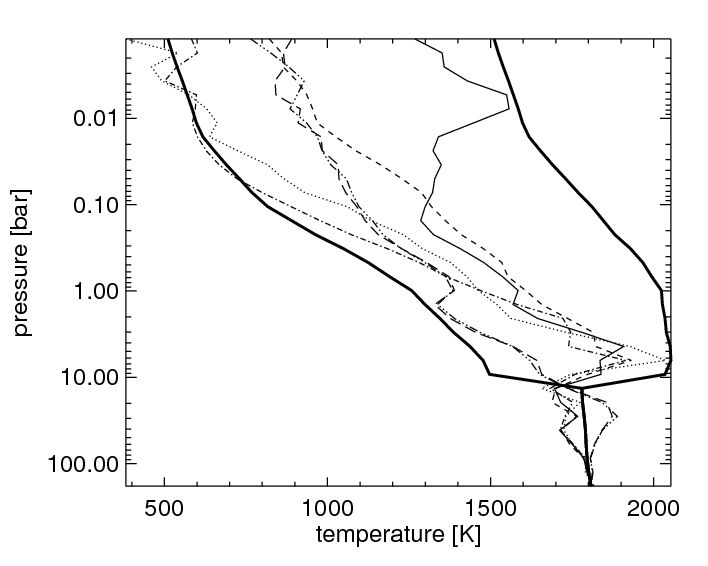}
\caption{Temperature-pressure profiles at six points around the
planet: the substellar point (\emph{thin solid}), the antistellar
point (\emph{dotted}), the equator at the east terminator
(\emph{dashed}), the equator at the west terminator
(\emph{dot-dashed}), the north pole (\emph{triple dot-dashed}), and
the south pole (\emph{long dashed}).  The thick solid lines show the
radiative relaxation (forcing) temperature profiles at the substellar
point and on the nightside above 10 bar, as well as the initial
temperature profiles below 10 bar (where there is no radiative forcing
by construction).  Note the presence of a dynamically-induced
temperature inversion (``stratosphere'') at the substellar point and
significant deviations from radiative equilibrium all around the
limb.}
\label{fig:tprof}
\end{center}
\end{figure}

In addition to analyzing the variation of flow pattern with height, it
is also worth considering the vertical temperature profiles around the
planet.  In Figure~\ref{fig:tprof} we plot the variation of
temperature with pressure for six representative locations: the sub-
and antistellar points, the equator at the east and west terminators,
and the north and south poles.  Also shown are the radiative
relaxation (forcing) profiles for the substellar point and everywhere
on the nightside.  Below 10 bar the atmosphere is not forced and
instead we show an initial temperature profile for that region.

Interestingly, the profile at the substellar point contains a
temperature inversion in the upper atmosphere.  Our forcing profile is
not inverted and so this inversion is due to \emph{dynamics alone}.
We note that this is not the first instance of a dynamically-induced
temperature inversion in hot Jupiter atmospheric models, with
inversions of similar magnitude found in \citet{CS06} and \citet{S08}.
However, given the growing observational evidence for temperature
inversions on some hot Jupiters
\citep{Harrington2007,Knutson2008,Machalek2008,Knutson2009b}, we find
it worthwhile to emphasize that dynamics could play a role in creating
such inversions.  Radiative transfer models used to explain these
observations generally invoke inversions that are much stronger than
the one presented here \citep{Burrows2008,Fortney2008}, but it is
possible that these models would not require as large an amount of
absorbers as they do \citep[e.g.,][]{Spiegel2009} if the role of
dynamics in creating such inversions were included.  \citet{Burrows2008} show that the presence of absorbing species are primarily responsible for the inversions, but advection of hot gas can play a secondary role.

Our terminator profiles at the east/west equator and the north/south
poles are significantly hotter than if they were in local radiative
equilibrium.  The forcing profile, which defines local radiative
equilibrium, is identical around the terminator and equal to the
nightside profile shown in Figure~\ref{fig:tprof}.  By comparison, we
find atmospheric temperatures at the terminator which are hotter than
the radiative equilibrium values by hundreds of Kelvin.  Although the
terminator relaxation profiles adopted in the present model cannot be
considered as very realistic, a systematic bias toward hotter
temperatures could have a significant impact on interpretations of
transit spectroscopy observations
\citep[e.g.,][]{Sing2008,Swain2008b}, especially since those
measurements probe regions above $\sim$100 mbar, where we find the
strongest deviations from radiative equilibrium.

Finally, an examination of vertical stability reveals that the
nightside temperature profile, at the antistellar point, is
convectively unstable across a few layers above the 40 mbar level.
All of the other plotted profiles are stably stratified.  Our code
does not include any special treatment for convective columns and the
occurrence of convection may thus introduce substantial errors in at
least part of the modeled upper atmosphere.  Convection acts as an
efficient vertical mixing process and its occurrence in the upper
atmosphere could facilitate the transport of minor absorbers, with
possible consequences for the radiative transfer and the overall
structure of the upper atmosphere \citep[e.g.,][]{Spiegel2009}.  An
adequate treatment of convective columns may thus be needed in future
hot Jupiter atmospheric models although we note that \citet{S09} do no
report any convective columns in more advanced models with explicit
non-gray radiative transfer.

\section{Discussion} \label{sec:discussion}

Our flow can be generally characterized as having an upper atmosphere
dominated by radiative forcing, a transition to an advection-dominated
regime at lower levels, and a low level of variability.  This agrees
at a basic level with the results of C\&S, but we find several
interesting differences that reveal possible areas of concern for
future modeling efforts.  We compare our results with C\&S in detail
in \S\ref{sec:cs} and discuss possible reasons for these differences.
In \S\ref{sec:vtransfer} and \S\ref{sec:jump} we emphasize concerns
related to vertical exchange between layers in the atmosphere and the
presence of a shock-like feature.

\subsection{Comparison with the work of Cooper \& Showman} \label{sec:cs}

It is reassuring that our results agree at a basic level with those of
C\&S since they both are the consequence of the same set of physical
equations applied to essentially identical conditions.  The pressure levels plotted in Figure 1 of \citet{CS05} correspond to the levels shown in our Figure \ref{fig:levels1}, top and bottom, and Figure \ref{fig:levels2}, bottom.  The top two levels show similar temperature and flow patterns, as well as matching maximum (supersonic) wind speeds.  Points of
interest are exactly where the models differ, as these indicate where
our methods of solution for these equations may break down in the hot
Jupiter regime.

First, however, we should focus on the greatest difference in set up
between our model and that of C\&S: the lower boundary.  The model of
C\&S reaches down to 3 kbar while we only descend to 220 bar.  This
may be the main cause of differences between our 20 bar level flow
(Figure~\ref{fig:levels2}, bottom panel) and that of \citet{CS05},
shown in their Figure 1c.  Our super-rotating equatorial jet descends
to $\sim$7 bar whereas theirs reaches below 50 bar \citep[compare our Figure~\ref{fig:uz} with][Fig. 2]{SC2006}.  In addition, their zonal profile shows a sharp distinction between the upper layers, where the flow is almost completely eastward, and the westward flow in the lower layers.  In contrast, we find flow in both directions in our active layers.  We partly attribute these differences in zonal flow to the extra momentum available from the deeper reservoir of
inert layers in C\&S.  Since the simulation begins with no winds, the net
angular momentum of the atmosphere must remain at the value set by the
bulk planetary rotation at all times.  For a deep eastward jet to
form, the positive excess momentum must be balanced by a westward flow
elsewhere in the atmosphere.  The deepest model levels develop such a
momentum balancing flow.  In support of this interpretation, we note
that our test run without any layers below 10 bar achieved a very
similar flow in the upper layers while the eastward jet was
constrained to terminate somewhat higher up in the atmosphere.  We
discuss the issue of vertical transport of momentum in more detail in
the next section (\ref{sec:vtransfer}), but here we simply note that
our fiducial model and the one without inert layers show good
agreement except in the lowest model levels, suggesting that the
difference in bottom boundary between our model and C\&S does not
strongly influence the modeled upper atmospheric flow.

Our next major difference with C\&S is that our model contains
significantly more detailed flow features, including vortices,
especially at levels below 100 mbar.  Since the large vortices in our
model tend to extend all the way up to the poles
(Figure~\ref{fig:levels1}, bottom panel) or flow around the poles
(Figure~\ref{fig:levels2}), their absence in the work of C\&S could be
related to the traditional difficulty in treating the polar regions in
a grid-based code.  In fact, we note that the more recent work by
\citet{S09}, which uses a cubed-sphere grid designed to address this
polar issue specifically, contain ``gyres'' that extend up to high
latitudes and may be analogous to the large scale vortices in our
model.

The dynamical scales for jets and vortices on hot Jupiters are
expected to be comparable to the planetary radius
\citep{SG02,Menou2003}.  \citet{S08} have argued that the horizontal
resolution in a model like ours or that in C\&S should be sufficient
to resolve the important atmospheric dynamical features.  The model
presented in \citet{S08} improves upon C\&S by using relaxation
temperature profiles calculated from one dimensional radiative
transfer models for 14 insolation angles away from the substellar
point, values of $\tau_{\mathrm{rad}}$ which depend on pressure and
temperature, and twice the horizontal resolution.  The flow pattern in
their model of HD 209458b (shown in their Figure 5) more closely
resembles our results than C\&S in several ways, especially regarding
the presence of a shock-like feature in the equatorial region around
longitude +145\degrees~E.  We note that runs at twice the horizontal  or vertical resolution produced very similar results, with only modest differences at the deepest model levels.

The emergence of sharp shock-like features in hot Jupiter atmospheres
may have a significant impact on the nature of the atmospheric flow,
as discussed in more detail in \S\ref{sec:jump}, but it may also lead
to noticeable differences between models using different methods of
solution, such as our model and that of C\&S.  Let us consider the
issue of hyperdissipation as a specific example. In order to model the
effect of dissipation on subgrid scales, atmospheric circulation
models include an artificial term that works to damp features at the
smallest resolved scales \citep[e.g.,][]{Stephenson94}.  C\&S make use
of a fourth-order hyperdiffusion term operating on the velocity field,
$dv/dt \propto -\kappa \nabla^4 v$, where $\kappa$ is a coefficient
adjusted to dissipate the smallest model scales on a timescale of
$\sim 30$ minutes.  On the other hand, eighth-order hyperdissipation
terms in our model act on the divergence and relative vorticity flow
fields, as represented in Equation~(\ref{eq:hyperdiss}), on a
timescale of $\sim 130$ minutes.  Both methods are meant to prevent
the build up of noisy features on small scales, but they may also have
the unintended consequence of diffusing small-scale shock-like
features that develop in hot Jupiter model flows.  Such distinction
could introduce an important algorithmic difference between models, to
the extent that applying a diffusion operator on the velocity or the
divergence field of a narrow divergent/convergent shock-like feature
is significant. While hyperdiffusion terms may not be treating such
features \emph{correctly}, as we discuss in \S\ref{sec:jump}, they may
also treat them \emph{differently} and this could cause some of the
dissimilarities between our model results and those of C\&S.

\subsection{Vertical exchange} \label{sec:vtransfer}

There is a discontinuity between active and inert layers in our model
and that of C\&S, with a sudden jump from long radiative timescales to
infinitely large ones.  A more realistic transition is used by
\citet{S08}, with radiative timescales continuously increasing down
through the atmosphere, so that there is no clear distinction between
active and inert layers.  Even then, the bottom boundary of the model
is located above the convective interior and the nature of the
interaction between the upper radiative atmosphere and the inner
convective planetary interior is left unspecified. Understanding the
nature of vertical exchanges in the deep atmospheres of gaseous giant
planets is of great interest, whether it concerns hot Jupiters or
Solar System giants \citep[e.g.,][]{LianShowman,Schneider}.

Vertical exchange of momentum and heat between our active and inert
layers is seen to occur, as discussed in \S\ref{sec:vertical}.  One
consequence of this interaction is apparent in Figure~\ref{fig:uz},
where the levels above 10 bar are dominated by an eastward flow (with
net excess angular momentum), with a compensating westward flow in the
deeper inert layers (with a net angular momentum deficit).  The
ability of the upper layers to acquire angular momentum from the lower
levels allows for stronger eastward flow than would otherwise be
possible, e.g., if the radiatively forced layers were isolated. In the
absence of momentum forcing, a much shallower atmosphere would have to
maintain this net angular momentum budget by balancing an eastward
flow with a westward flow essentially within the same atmospheric
layer.  This situation appears to be realized in the shallow hot
Jupiter model presented by \citet{MR08}, leading to the development of
strong horizontal shear and barotropic instabilities.  By contrast,
vertical transport of angular momentum between layers can occur
throughout a deep atmosphere. It connects the dynamics of the upper
atmosphere with various interior processes, with possibly significant
implications for a planet's tidal evolution and cooling history
\citep[e.g.,][]{GS02,SG02}.

When modeling the upper atmosphere, the main difficulty in correctly
treating interaction with the deeper layers and the planetary interior
comes from our ignorance of the physical processes involved, as well
as issues of timescales.  In the upper atmosphere, the timescales of
interest are much shorter than those in the interior.  In fact, layers
below 10 bar in our model have not reached a stationary state by the
end of the run, as shown in Figure~\ref{fig:ke}.  They continue to
acquire kinetic energy from interaction with the radiatively forced
region, although this does not seem to significantly affect the upper
atmospheric flow itself.  As mentioned earlier, we ran a test model
excluding inert layers, i.e., with a model lower boundary set at 10
bar, and found significant differences only in the bottom few layers,
with a stronger westward flow just above 10 bar.  This suggests that
the nature of the upper atmospheric flow is dominated by radiative
forcing rather than by exchange with deeper layers but the issue of
vertical exchange of heat and momentum in the deep atmospheres of hot
Jupiters probably deserves further attention in the future.

\subsection{The shock-like feature}\label{sec:jump}

\begin{figure}[ht]
\begin{center}
\includegraphics[width=0.7\textwidth]{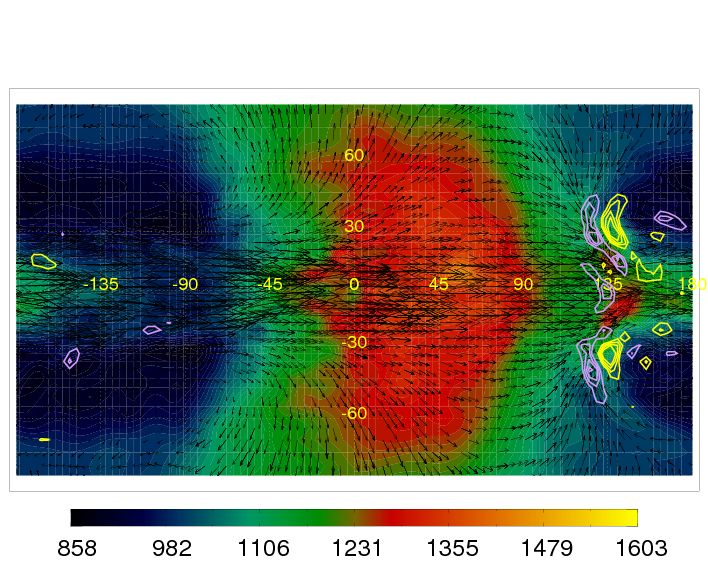}
\caption{Contours of horizontal flow convergence (in lavender) and
divergence (in yellow) plotted over the temperature (in K) and wind
map at the 150 mbar pressure level.  The convergence/divergence
contour levels are set at $\pm1\times 10^{-4}$, $1.5\times 10^{-4}$,
$2\times 10^{-4}$, and $2.5\times 10^{-4}$ s$^{-1}$.  Clearly, the
bulk of the flow has $|\nabla \cdot \mathbf{v}|<1\times 10^{-4}$
s$^{-1}$.  The narrow region of strong convergence around longitude
+135\degrees~is what we refer to as a shock-like feature. An increase
in temperature around longitude +135\degrees~seen in the bottom panel
of Figure~\ref{fig:levels1}, just one model level below the one shown
here, is consistent with adiabatic heating by a downward flow
associated with this shock-like feature.}
\label{fig:jump}
\end{center}
\end{figure}

The bottom panel of Figure~\ref{fig:levels1} shows the presence of a
chevron-shaped feature in the 220 mbar temperature map, at longitude
+135\degrees~E.  In Figure~\ref{fig:jump}, we over-plot contours of
horizontal flow convergence (and divergence) on a temperature and wind
map one level higher in the atmosphere, at 150 mbar. There is a narrow
region of strong horizontal flow convergence at the 150 mbar level
which matches well the temperature feature seen at the same longitude,
at the 220 mbar level, in Figure~\ref{fig:levels1}.  We identify this
flow feature as ``shock-like'' because unusually strong local flow
convergence is a signature of hydrodynamical shocks. Our numerical
solver does not include any special treatment of the shock physics,
however.  The feature extends vertically across a range of layers,
with hints of it even as high as 2.5 mbar (Figure~\ref{fig:levels1},
top panel), but it is most apparent in levels from 50 - 150 mbar.  The
regions of strong horizontal convergence generally coincide with areas
of higher temperatures for the atmospheric gas. Although we have not
attempted a more detailed characterization of this flow feature, we
note that it is consistent with adiabatic heating of a downward flow
associated with horizontal convergence above \citep[as discussed
by][]{S09}.  The influence of this localized source of heating may
extend rather deep into the atmosphere, so that the increased
temperature region centered around longitude +135\degrees~at the 4.4
bar level (Figure~\ref{fig:levels2}, top panel) may also be a
consequence of the flow higher up, rather than being caused by
horizontal advection of heat away from the substellar point.

Recently, \citet{S09} reported a similar feature in their model of the
hot Jupiter HD 189733b, which they identified as a hydraulic
jump\footnote{A hydraulic jump is the equivalent of a shock for an
incompressible flow \citep[e.g.,][]{Kundu}.}. Another such feature can
be seen in their model for HD 209458b.  In fact, a similar feature
also seems to be present, but not commented upon, in the models of
these same planets by \citet{S08}.  We note that the primitive
equations of meteorology impose hydrostatic balance, which filters out
vertically propagating sound waves.  Horizontal sound waves, however,
are permitted given appropriate boundary conditions
\citep{Kalnay,Vallis2006}, so that finite amplitude waves of this
nature (also known as Lamb waves) could presumably steepen into
shocks.  The partial-globe model of \citet{DobbsDixon2008} solves the
full three dimensional Navier-Stokes equations and these authors also
find a shock-like, chevron-shaped feature in their hot Jupiter model
around +135\degrees~longitude (see their Figure 6).  Since shock-like
features are present in models of increasing complexity \citep[this
work,][]{S08,S09}, in a model that includes fully three dimensional
sound waves \citep{DobbsDixon2008}, and in models for different
planets, it may be that the development of such features is a natural
outcome of the hot Jupiter circulation regime. While the steepening of
finite amplitude sound (or Lamb) waves is a possible path to the
formation of shocks in an atmosphere with transonic wind speeds, the
detailed nature of these shock-like features remains to be
investigated and understood.

Most meteorological models are not designed to treat the three
dimensional compressible nature of atmospheric flows or the formation
of shocks because these are issues which are clearly subdominant in
the context of Earth atmospheric modeling, with very subsonic wind
speeds.  Any model that assumes hydrostatic balance has filtered out
vertically propagating sound waves, while the use of rigid top and
bottom boundary conditions would exclude even horizontally propagating
sound waves \citep[i.e., Lamb waves;][]{Kalnay}. In meteorological
models that do admit horizontal sound waves, there usually is no
explicit treatment of shock physics, so that shocks would be subject
to ill-constrained conditions on the smallest resolved model scales
(see also the discussion of hyperdiffusion in \S
\ref{sec:cs}). Physically, the dissipation of bulk kinetic energy in a
shock, and its conversion to localized heating, could have a
substantial impact on the flow, and one that cannot be easily
estimated without an explicit treatment of shocks.
\citet{Goodman2009} clearly emphasizes various concerns regarding this
issue.  Future hot Jupiter models may thus need to include
shock-capturing schemes to more faithfully describe this aspect of
their new circulation regime.

\section{Summary and Conclusions} \label{sec:summary}

It is reassuring to find a basic level of consistency between our work
and that of C\&S.  In fact, the general agreement between our hot
Jupiter model and more complex versions described by \citet{S08,S09}
seems to indicate that some of the basic attributes of the circulation
regime in hot Jupiter atmospheres has been successfully captured.
Nevertheless, significant discrepancies also emerge from our direct
comparison with the work of C\&S. These discrepancies could have their
origin in the different methods of solution adopted and/or the fact the
equations solved are missing some of the physics at work in the hot
Jupiter context.

Meteorological solvers like our pseudo-spectral IGCM and the
grid-based code used by C\&S have been successfully compared to each
other in the context of Earth atmospheric modeling \citep{HS94}. They
very accurately solve the equations of meteorology and produce fully
consistent results, which are also in agreement with the known general
circulation regime on Earth.  In the hot Jupiter modeling context, we
cannot be certain of the accuracy and validity of these models until
they can likewise produce consistent results.  The discrepancies
identified here between our model results and those of C\&S suggest
that one or both numerical solvers do not faithfully capture all of
the atmospheric physics at work in the hot Jupiter context.  Areas of
specific concern for future hot Jupiter atmospheric modeling include
the following.

\begin{itemize}

\item The interaction between the radiatively forced upper atmosphere,
deeper atmospheric layers, and the convective planetary interior.
Although the upper atmosphere seems to be rather weakly sensitive to
the lower boundary condition on the relatively short timescales
described by our model (hundreds of planet days), the nature of heat
and momentum exchanges between the modeled atmosphere and the
unmodeled planetary interior could be a significant source of
uncertainties for hot Jupiter models.

\item Dissipation of bulk kinetic energy and conversion to localized
heating in shock-like features.  Explicit treatment of this energy
conversion process may have a first order effect on the flow
thermodynamics, as emphasized by \citet{Goodman2009}.  Previous
atmospheric models, in circulation regimes with typically subsonic
wind speeds, did not have to treat such shock-like features.

\item Convection.  If atmospheric columns are found to be convective
in hot Jupiter models with more advanced treatments of radiative
transfer and atmospheric chemistry, it will be necessary to implement
convective adjustment schemes to properly describe the rapid vertical
mixing of entropy and momentum. Vertical mixing of radiatively active
species could also have important consequences for the atmospheric
structure.
\end{itemize}

Finally, our hot Jupiter model results also inform some of the current
efforts focused on the interpretation of a growing set of hot Jupiter
observational data.  We have emphasized the possibility that
atmospheric dynamics alone could produce upper atmospheric temperature
inversions, albeit not necessarily of the strength typically required
to explain observed hot Jupiter secondary eclipse spectra.
Nevertheless, it may be important to recognize the role of dynamics in
this phenomenon.  In addition, we have found that temperature profiles
along the terminator in our model upper atmosphere are systematically
raised above the local radiative equilibrium values.  This is a
potential source of bias for interpretations of transmission spectra
which often rely on the assumption of temperature profiles in
radiative equilibrium.

\acknowledgements

We thank Adam Burrows, James Cho, Lorenzo Polvani, and Adam Showman for useful
discussions.  This work was supported by the NASA OSS program,
contract \#NNG06GF55G, a NASA Graduate Student Research Program
Fellowship, contract \#NNX08AT35H, and the Spitzer Space Telescope
Program, contract \# JPLCIT 1366188. We also thank Drake Deming for
supporting this work.

\end{document}